\documentclass[journal]{IEEEtran}
\usepackage{color}
\usepackage{glossaries}
\usepackage[printonlyused]{acronym} 
\usepackage{amsmath}
\usepackage[caption=false,font=footnotesize]{subfig}
\usepackage{bm,amssymb}
\usepackage{amsmath,amssymb,amsfonts,amsthm}
\usepackage{graphicx}
\usepackage{algorithm}
\usepackage{algpseudocode}

\usepackage{tikz}
\usepackage{balance}
\usepackage[utf8]{inputenc}
\usepackage{pgfplots} 
\usepackage{pgfgantt}
\usepackage{pdflscape}
\pgfplotsset{plot coordinates/math parser=false}

\usepackage{graphicx}
\usepackage[noadjust]{cite}
\usepackage{pifont}
\usepackage{bm}

\theoremstyle{plain}
\DeclareMathAlphabet\mathbfcal{OMS}{cmsy}{b}{n} 
\usepackage{tcolorbox}
\usepackage{enumitem}
\usepackage{graphicx}
\usepackage[normalem]{ulem}
\usepackage{multirow}

\newcommand\numberthis{\addtocounter{equation}{1}\tag{\theequation}}

\acrodef{GPS}{Global Positioning System}
\acrodef{LOS}{line-of-sight}
\acrodef{NLOS}{non-line-of-sight}
\pgfplotsset{compat=1.17} 

\begin{document}
 
\title{An Iterative 5G Positioning and Synchronization Algorithm in NLOS Environments with Multi-Bounce Paths}

 \author{Zhixing Li, Fan~Jiang,~\IEEEmembership{Member,~IEEE,} ~Henk~Wymeersch,~\IEEEmembership{Senior~Member,~IEEE}, Fuxi~Wen,~\IEEEmembership{Senior~Member,~IEEE} 
 \thanks{Z. Li and F. Wen are with the School of Vehicle and Mobility, Tsinghua University, Beijing, China, Email: wenfuxi@tsinghua.edu.cn.} %
  \thanks{F. Jiang and H. Wymeersch are with the Department of Electrical Engineering of Chalmers University of Technology, Gothenburg, Sweden. Email: henkw@chalmers.se.}
  }
\markboth{IEEE Short paper \today}%
{Shell \MakeLowercase{\textit{et al.}}: Bare Demo of IEEEtran.cls for IEEE Journals}
\maketitle

\begin{abstract}
5G positioning is a very promising area that presents many opportunities and challenges. Many existing techniques rely on multiple anchor nodes and line-of-sight (LOS) paths, or single reference node and single-bounce non-LOS (NLOS) paths. However, in dense multipath environments, identifying the LOS or single-bounce assumptions is challenging.
The multi-bounce paths will make the positioning accuracy deteriorate significantly.
We propose a robust 5G positioning algorithm in NLOS multipath environments.
The corresponding positioning problem is formulated as an iterative and weighted least squares problem, and different weights are utilized to mitigate the effects of multi-bounce paths. Numerical simulations are carried out to evaluate the performance of the proposed algorithm. Compared with the benchmark positioning algorithms only using the single-bounce paths, similar positioning accuracy is achieved for the proposed algorithm.  
\end{abstract}


\IEEEpeerreviewmaketitle


\section{Introduction}

5G New Radio offers great opportunities for accurate localization by introducing large bandwidth, high carrier frequency, and large antenna array.
Most of the state-of-the-art localization techniques are designed based on multiple anchor nodes and line-of-sight (LOS) paths, or single reference node
and single bounce non-LOS (NLOS) paths radio propagation  \cite{Xiao2021}. 
A low complexity, search-free 5G mmWave localization and mapping method that is able to operate using single-bounce diffuse multipath is proposed in \cite{Wen2021}, where LOS and specular multipath are not required. In \cite{Shikur2014}, the authors propose
a localization algorithm for use in NLOS environments. The single bounce scattering model is utilized to model the NLOS propagation and to estimate the position of a mobile station when the observations are the time-difference-of-arrival (TDOA), the angle-of-departure (AOD), and the angle-of-arrival (AOA). The proposed algorithm uses the underlying geometry of the radio propagation paths to estimate the position of the mobile station.
Paper \cite{Wei2011} focuses on indoor scenarios which are multipath and rich scattering environments considering NLOS propagation. Based on the measured AOD, AOA, and time-of-arrival (TOA), a three dimensional (3D) least squares (LS) positioning algorithm is proposed assuming a single-bounce reflection in each NLOS propagation path.

However, for 5G positioning in
dense multipath environments, the LOS or single 
bounce assumptions can be invalid. The multi-bounce paths
will make the positioning accuracy deteriorate significantly. 
One option is to remove paths directly based on geometric grounds if they are not LOS or single-bounce, using the angle difference between the LOS path and a possible multi-bounce path \cite{Kakkavas2021}. 
Because the channel gains for the multiple-bounce paths are much smaller than that of the LOS and single-bounce NLOS paths, a part of previous research ignores the multi-bounce paths or just uses received power to identify these multi-bounce paths \cite{mao2020novel}.
However, the results in \cite{Ko2021} show that it could be difficult to distinguish the single-bounce and double-bounce paths from the multi-bounce paths by using the received power only. 
Furthermore, the study in \cite{geng2021joint} indicates that some specific spatial shape and the material of surface will also influence the identification. It is shown in \cite{9769384} that multi-bounce paths should be considered in real environment because the power and the total number of multi-bounce paths occupy a large proportion.

In this paper, we investigate robust positioning techniques to relief these LOS and single-bounce  
assumptions using the available channel parameter measurements, such as AOD, AOA, TOA, and channel gain \cite{Ge2020}.
The main contributions are summarized as follows:
\begin{itemize}
\item A weighted least square (WLS) based robust 5G positioning and synchronization algorithm based on single BS in NLOS multipath environment is proposed. In the proposed WLS-based algorithm, different weights are utilized to mitigate the effects of multiple bounce paths.

\item Based on the generalized likelihood ratio test (GLRT) method, we propose an iterative strategy to distinguish single-bounce and multi-bounce paths.

\item A numerical study on the distribution of the measurement errors is conducted, which demonstrates that the proposed algorithm achieves robust localization performance.
\end{itemize}
 
\section{Problem Formulation}
\label{sec2}

\begin{figure}
\centering
\includegraphics[width=0.45\textwidth]{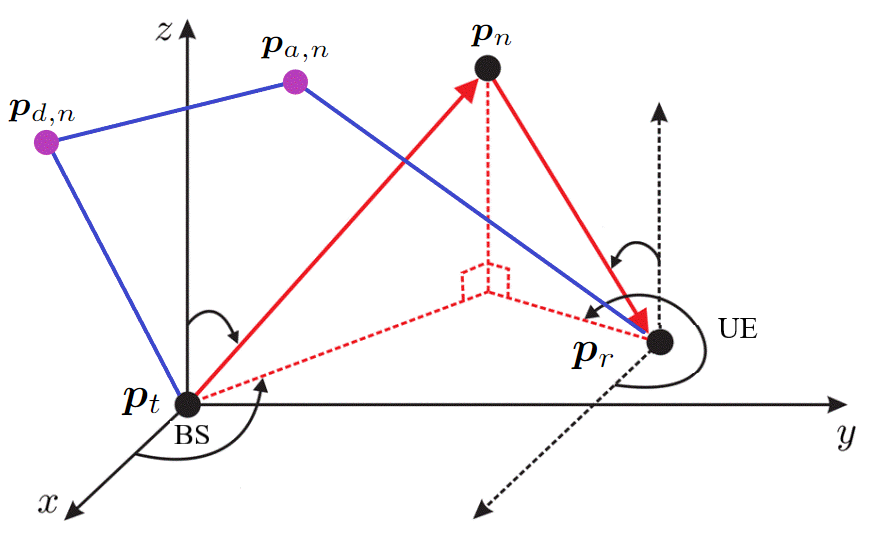}
\caption{System model with single-bounce and multiple-bounce NLOS propagation paths, the user equipment “UE”, the base station “BS”, and the incidence points $\bm{p}_{d,n}$ and $\bm{p}_{a,n}$ for multiple bounce paths, $\bm{p}_n = \bm{p}_{d,n}=\bm{p}_{a,n}$ for single bounce paths. For the LOS path, we introduce $\bm{p}_{0}$ as any point on the line segment strictly excluding $\bm{p}_{\rm{t}}$ and $\bm{p}_{\rm{r}}$.} 
\label{fig:scenario}
\end{figure}

We consider a down-link 3D positioning scenario 
with a single base station (BS) with known location $\bm{p}_{t} = [x_{t},y_{t},z_{t}]^{\mathsf{T}}$ and single user equipment (UE) with unknown location  $\bm{p}_{r} = [x_{r},y_{r},z_{r}]^{\mathsf{T}}$ and clock bias $\tau_B$. We assume that the orientation between BS and UE is known. As shown in Fig.~\ref{fig:scenario}, the complex propagation environment leads to single-bounce NLOS paths (show in red) and multi-bounds NLOS paths (shown in blue), in addition to a possible LOS path (not shown). 

Based on a channel parameter estimation  method, we obtain, for each path $n$, estimates of the channel gain (amplitude) $\gamma_{n}\ge 0$, 
 the azimuth and elevation angles of AOD,  denoted by $(\phi_{d,n}, \theta_{d,n})$; the azimuth and elevation angles of AOA, denoted by  $(\phi_{a,n}, \theta_{a,n})$, the TOA $\tau_{n}=d_n/c+\tau_B$, where $d_n$ is the total propagation distance, $c$ is the speed of light, and 
 $\tau_B$ is the unknown clock bias caused by imperfect synchronization between BS and UE. For each path, it is unknown whether it is LOS, single-bounce, or multiple-bounce.

We make use of the following essential geometric relations, which hold for LOS and single-/double-bounce paths, but not for multi-bounce paths larger than two \cite{Wen2021}: 
\begin{subequations}
\begin{align}
    d_{n}& = \Vert\bm{p}_{d,n}-\bm{p}_{t}\Vert + \Vert\bm{p}_{a,n}-\bm{p}_{d,n}\Vert +\Vert\bm{p}_{a,n}-\bm{p}_{r}\Vert\\
    \phi_{a,n}& =\pi+\mathrm{atan}2\left(y_{a,n}-y_{r},x_{a,n}-x_{r}\right)\\
    \theta_{a,n}&=\mathrm{asin} \left( \frac{ z_{a,n}-z_{r} }{\|\bm{p}_{a,n}-\bm{p}_{r}\|} \right)\\
    \phi_{d,n}&=\mathrm{atan}2\left(y_{d,n}-y_{t}, x_{d,n} - x_{t}\right)\\
    \theta_{d,n}&=\mathrm{asin} \left(\frac{ z_{d,n}-z_{t}}{\|\bm{p}_{d,n}-\bm{p}_{t}\|} \right).
\end{align}
\end{subequations}
where $\| \cdot \|$ is Euclidean norm.
Our goal is to estimate the UE location $\bm{p}_{r}$, based on the estimated channel parameters.  We tackle the problem based on the methods from \cite{Wen2021}.

\section{Proposed Method}
\label{sec3}

In order to solve the positioning problem, we first establish identities that hold for each path $n$, be it LOS, single-bounce, or multi-bounce. Then we describe a method that can estimate the UE position and clock bias from at least 2 multipath channel parameter estimates (the 2 multipath should be either LOS or single-bounce paths). Finally, we use both these results to propose our final method, which involves using a set of ordered paths, combined with change detection in the positioning residuals. 
\subsection{Identities for 5G Positioning and Synchronization}
Before describing the proposed method, we first list identities valid for any path $n$, be it LOS, single-bounce, or multi-bounce. We first define
\begin{equation}\label{eqDir}
    \bm{f}_{t,n}=\begin{bmatrix}
       \cos(\hat{\theta}_{d,n})\cos(\hat{\phi}_{d,n}) \\ \cos(\hat{\theta}_{d,n})\sin(\hat{\phi}_{d,n}) \\ \sin(\hat{\theta}_{d,n})
     \end{bmatrix},
\end{equation}
which points along the AOD of path $n \in \{1, 2, \ldots,{N}\}$; and $\mathbf{f}_{r,n}$ is defined equivalently for the AOA, pointing from the UE towards the $n$-th artificial specular point $\bm{p}_{a,n}$: 
\begin{equation}\label{eqDir2}
    \bm{f}_{r,n}=\begin{bmatrix}
       \cos(\hat{\theta}_{a,n})\cos(\hat{\phi}_{a,n}) \\ \cos(\hat{\theta}_{a,n})\sin(\hat{\phi}_{a,n}) \\ \sin(-\hat{\theta}_{a,n}).
     \end{bmatrix}.
\end{equation}
Then, we have the following relations, valid for double bounce, single bounce and
direct paths:
\begin{equation}
\label{eq-double-bounce}
 \left\{
\begin{aligned}
\bm{p}_{d,n} &= \bm{p}_{t} + \xi_{d,n} d_n \bm{f}_{t,n}\\
\bm{p}_{a,n} &= \bm{p}_{r} + \xi_{a,n} d_n \bm{f}_{r,n}\\
\xi_{d,n} &+ \xi_{a,n} \leq 1, \text{ and } 0 < \xi_{d,n},\xi_{a,n} < 1
\end{aligned}
 \right.
 \end{equation}
where
$\xi_{d,n}$ and $\xi_{a,n}$ are unknown and represent the fraction of the delay $\hat{\tau}_n$ that is attributed to the line from BS to the first scatter point $\bm{p}_{d,n}$ and from UE to the second scatter point  $\bm{p}_{a,n}$. Note that  of $\xi_{d,n} + \xi_{a,n} = 1, \text{ and } 0 < \xi_{d,n},\xi_{a,n} < 1$ for single bounce paths and the LOS path. We now introduce 
$\bm{e}_n = \bm{p}_{d,n} - \bm{p}_{a,n}$,
then we can express \eqref{eq-double-bounce} as 
\begin{align}
\label{eq-general}
    \bm{p}_r = \bm{p}_t + \mathbf{e}_n + \left(\xi_{d,n} \mathbf{q}_{t,n} -  \xi_{a,n} \mathbf{q}_{r,n}\right), \forall n
\end{align}
where $\mathbf{q}_{t,n} = d_n\bm{f}_{t,n}$ and $\mathbf{q}_{r,n} = d_n\bm{f}_{r,n}$.
For multi-bounce paths, we have $\| \bm{e}_n \|_2 > 0$.
Because $\mathbf{e}_n$ is a unknown variable, the range of the feasible solutions for \eqref{eq-general} is unbounded. Therefore, it is challenging to estimate the UE position by solving a set of linear equations using WLS methods.

\subsection{WLS-based 5G Positioning and Synchronization}

For LOS and single-bounce cases, we have $\mathbf{e}_n = \mathbf{0}$, and $ \xi_{a,n} = 1-\xi_{d,n} $, 
then (\ref{eq-general}) can be simplified as
\begin{align}
    \bm{p}_{r} = 
 \bm{p}_{t} +  \xi_{d,n} \mathbf{q}_{t,n} - (1-\xi_{d,n}) \mathbf{q}_{r,n},\label{eq:basicRelation}
\end{align}
The UE position can be determined if there are multiple single-bounce paths. Specifically, from (\ref{eq:basicRelation}), we establish 
\begin{align*}
\boldsymbol{p}_{r} & = \boldsymbol{p}_{t} + c\left(\hat{\tau}_n - \tau_B \right) \xi_{d,n} \boldsymbol{f}_{t, n} - c \left(\hat{\tau}_n - \tau_B \right)  \left( 1 - \xi_{d,n}\right) \boldsymbol{f}_{r, n} \\
& = \boldsymbol{p}_{t} - c \hat{\tau}_n \boldsymbol{f}_{r, n} + c \tau_B \boldsymbol{f}_{r, n} + c\xi_{d,n}\left( \hat{\tau}_n - \tau_B \right) \left( \boldsymbol{f}_{t, n} + \boldsymbol{f}_{r, n}\right) \\
& = {\boldsymbol{\delta}}_n + c \tau_B \boldsymbol{f}_{r, n} + {\xi}_{d,n} {\boldsymbol{u}}_n - c \tau_B {\xi}_{d,n} \left( \boldsymbol{f}_{t, n} + \boldsymbol{f}_{r, n}\right) , \label{eqPrSynErr} \numberthis
\end{align*}
where ${\boldsymbol{\delta}}_n = \boldsymbol{p}_{t} - c \hat{\tau}_n \boldsymbol{f}_{r, n}$, ${\boldsymbol{u}}_n = c \hat{\tau}_n \left( \boldsymbol{f}_{t, n} + \boldsymbol{f}_{r, n}\right)$, ${\boldsymbol{v}}_n = c  \left( \boldsymbol{f}_{t, n} + \boldsymbol{f}_{r, n} \right)$, and $\tau_{\xi,n} = \tau_B {\xi}_{d,n}$. 
We further rewrite \eqref{eqPrSynErr} as
\begin{align}
\begin{bmatrix} \boldsymbol{I}_{3} &  - \boldsymbol{u}_n & -c \boldsymbol{f}_{r, n} &  \boldsymbol{v}_n \end{bmatrix} \begin{bmatrix} \boldsymbol{p}_{r} \\  {\xi}_{d,n} \\ \tau_B \\ \tau_{\xi,n} \end{bmatrix} = \boldsymbol{\delta}_n.
\end{align}
 With the estimations of $N$ sets of multipath channel parameters, we can establish $3N$ linear equations with $(4$+$2N)$ unknowns
 $ \bm{\mu} = \begin{bmatrix} \boldsymbol{p}^{\mathrm{T}}_{r},\,  {\xi}_{d,1},\, \cdots, \,  {\xi}_{d,N} ,\, \tau_B, \tau_{\xi,1}, \cdots, \tau_{\xi,N} \end{bmatrix}^{\mathrm{T}} $. 
 Therefore, with $N\ge 2$ multipath components,
 we have
\begin{equation}
\label{eq11}
   \boldsymbol{U} \bm{\mu} =  \boldsymbol{\delta}
\end{equation}
where $\boldsymbol{\delta} = \begin{bmatrix}   
\boldsymbol{\delta}_1^{\mathrm{T}}, & \boldsymbol{\delta}_2^{\mathrm{T}}, & \cdots & \boldsymbol{\delta}_N^{\mathrm{T}}
\end{bmatrix}^{\mathrm{T}} \in \mathbb{R}^{3N\times 1}$
and $\boldsymbol{U}\in \mathbb{C}^{3N \times (2N+4)}$ is defined as
\begin{align*}
\boldsymbol{U} = \left[ 
\begin{array}{cccccccc}
\boldsymbol{I}_{3}  & -\boldsymbol{u}_1  & & \bm{0} & -c \bm{f}_{r,1} & \boldsymbol{v}_1  & & \bm{0}\\
\vdots  &  & \ddots &  &  \vdots &    & \ddots & \\
\boldsymbol{I}_{3}  & \bm{0} & & -\boldsymbol{u}_N & -c \bm{f}_{r,1} & \bm{0}  & &  \boldsymbol{v}_N
\end{array}
\right]. \numberthis
\end{align*}
The variable ${\bm{\mu}}$ can be estimated with a weighted least-square solution as
\begin{align}
\hat{\bm{\mu}} & =  \left( \boldsymbol{U}^{\mathrm{H}} \mathbf{W} \boldsymbol{U} \right)^{-1} \boldsymbol{U}^{\mathrm{H}} \mathbf{W} \boldsymbol{\delta}, \label{eq_esty}
\end{align}
where the block diagonal matrix 
\begin{align*}
\mathbf{W} =  \text{blkdiag} \big[ w_1\boldsymbol{I}_3,\,   w_2\boldsymbol{I}_3,  \cdots,  w_N\boldsymbol{I}_3 \big] \in \mathbb{R}^{3N \times 3N},
\end{align*}
accounts for the normalized weight of each path, via $w_n = \gamma_n/(\sum_{n} \gamma_n). $  
Note that $\mathbf{W} = \boldsymbol{I}_{3N}$ is the conventional LS solution. Finally, the estimated UE position is $\hat{\bm{p}}_r = \hat{\bm{{\mu}}}_{[1:3]}$

\subsection{WLS with Change Detection}
\label{sec3b}

 We first order the paths, e.g., based on delay (from smallest to largest), or based on amplitude (from largest to smallest). Generally speaking, the first two arrival paths are usually LOS or single-bounce paths, because most multi-bounce paths have larger TOA.
 We thus use the first $k = 2$ paths in \eqref{eq_esty} to determine an initial estimate, say $\bm{p}_r^{(1)}$.
Similarly, we compute $\bm{p}_r^{(t)}$ from the first $k+t-1$ paths, $t=2, 3, \ldots$, using \eqref{eq_esty}. 
 \begin{table}
 \caption{Relative UE estimation error using different number of the first coming paths.}
 \centering
\begin{tabular}{|c |c|c|c|}
\hline
Step $t$ & Number of paths & Estimated $\bm{p}_r$ & Relative estimation error \\ \hline
1 & 2    & $\bm{p}_r^{(1)}$  &  0 \\ \hline
2 & 3                                & $\bm{p}_r^{(2)}$  & $\Delta_1 = \| \bm{p}_r^{(2)} - \bm{p}_r^{(1)} \|$       \\ \hline
3 & 4                                & $\bm{p}_r^{(3)}$ & $\Delta_2 = \| \bm{p}_r^{(3)} - \bm{p}_r^{(1)} \|$        \\ \hline
4 & 5                                & $\bm{p}_r^{(4)}$ & $\Delta_3 = \| \bm{p}_r^{(4)} - \bm{p}_r^{(1)} \|$   \\ \hline
\end{tabular}
\label{table1}
\end{table}
From these estimates, we compute the instant relative estimation error, 
\begin{equation}
\label{eq16delta}
 \Delta_{t} = \|\bm{p}_r^{(t)} - \bm{p}_r^{(1)} \|, 
\end{equation}
and constructing the following positioning residual vector (see also Table \ref{table1})
\begin{equation}
\label{eq17ot}
\bm{\Delta}^{(t)} = \begin{bmatrix}
      \Delta_1 & \Delta_2 & \cdots & \Delta_t
     \end{bmatrix}^{\mathrm{T}}.
\end{equation}
From (\ref{eq-general}), we recall that for multi-bounce paths, $\| \bm{e}_n \| > 0$, 
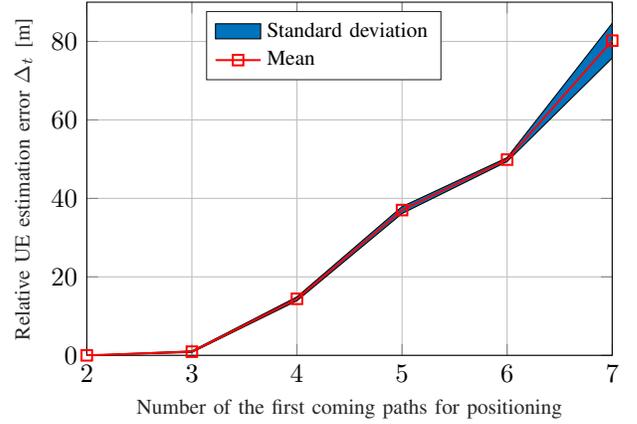
\begin{figure}[ht] 
\centering
%
%
\definecolor{mycolor1}{rgb}{0.00000,0.45000,0.74000}%
\begin{tikzpicture}

\begin{axis}[%
width=2.75in,
height=1.85in,
at={(0.758in,0.481in)},
scale only axis,
xmin=2,
xmax=7,
xlabel style={font=\color{white!15!black}},
xlabel={\footnotesize{Number of the first coming paths for positioning}},
ymin=0,
ymax=90,
ylabel style={font=\color{white!15!black}},
ylabel={\footnotesize{Relative UE estimation error $\Delta_t$} [m]},
axis background/.style={fill=white},
xmajorgrids,
ymajorgrids,
legend style={at={(0.227,0.779)}, anchor=south west, legend cell align=left, align=left, draw=white!15!black,font=\footnotesize}
]

\addplot[area legend, draw=black, fill=mycolor1]
table[row sep=crcr] {%
x	y\\
2	0\\
3	1.12938893006427\\
4	14.9047967379429\\
5	37.9006148408841\\
6	50.3846956803978\\
7	84.6219426623374\\
7	75.7704092647887\\
6	49.3830110014926\\
5	36.142843979686\\
4	13.9100886160752\\
3	0.742293810882616\\
2	0\\
}--cycle;
\addlegendentry{Standard deviation}

\addplot [color=red, line width=0.75pt, mark=square, mark options={solid, red}]
  table[row sep=crcr]{%
2	0\\
3	0.935841370473442\\
4	14.407442677009\\
5	37.0217294102851\\
6	49.8838533409452\\
7	80.196175963563\\
};
\addlegendentry{Mean}

\end{axis}
 
\end{tikzpicture}%
\caption{Motivation of using the slope change detection method to selected the  single bounce paths. The first 3 paths are single-bounce paths, while the multiple bounce paths appear starting from path 4.}
\label{fig2:RelativeError}
\end{figure}
so that for the proposed estimator in \eqref{eq_esty}, as shown in Fig. \ref{fig2:RelativeError}, we expect a larger relative UE estimation error $\Delta_t$ when a multi-bounce path is included for WLS estimation. Since paths later in the ordering are more likely to be multi-bounce,\footnote{This statement will be corroborated in the numerical results.} we can interpret this as 
a single-sensor 
change detection problem with observations  $\bm{\Delta}^{(t)}$.

\subsubsection{Change Point Detection}
Change point detection is an active research area in statistics due to its importance across a wide range of applications. The change-point can be modeled as a shift in the means of the observations, which is good for modeling an abrupt change \cite{Killick2012}. However, in many applications, the change point may cause a gradual change to the observations, which can be well approximated by
a slope change in the means of the observations \cite{Xie2018}.
Under the hypothesis of no change, the observations are drawn from $\mathcal{N}\left(\mu, \sigma^{2}\right)$, i.e., with 
a  fixed mean $\mu$ and variance $\sigma^2$. When a change occurs at  
 $\kappa$ (the unknown change-point), then
the mean of the observations changes
linearly from the change-point time $\kappa + 1$, which is given by $\mu + s (t-\kappa)$
for all $t > \kappa$, and the variance remains $\sigma^2$. Here, the unknown rate of change  is $s \neq 0$.
The above setting can formulate as the following hypothesis testing problem:
\begin{equation}
\begin{aligned}
\mathcal{H}_{0}: & \Delta_{i} \sim \mathcal{N}\left(\mu, \sigma^{2}\right), i\ge 1 \\
\mathcal{H}_{1}: & \Delta_{i} \sim 
\begin{cases}
\mathcal{N}\left(\mu, \sigma^{2}\right)&  i\le \kappa \\
\mathcal{N}\left(\mu+s(i-\kappa), \sigma^{2}\right)&  i>\kappa.
\end{cases}
\end{aligned}
\end{equation}
 
Our goal is now to establish a detection rule that detects as soon as possible after a change-point occurs and avoid raising false alarms when there is no change.
It can be solved efficiently by generalized likelihood ratio test (GLRT) method \cite{Besson2017}. 
Since the observations are independent, for an assumed change-point location
$\kappa = k$, the log-likelihood for observations up to time $t > k$ is given by \cite{Xie2018}
\begin{equation}
\label{eqll20}
\ell_{k, t, s} = \frac{1}{2 \sigma^{2}} \sum_{i=k+1}^{t}\left[2 s\left(\Delta_{ i}-\mu\right)(i-k)-s^{2}(i-k)^{2}\right].
\end{equation}

The unknown rate-of-change $s$ can be replaced by its maximum likelihood estimator.
Given the current number of observations $t$ and a assumed change-point location $k$, by setting the derivative of the log-likelihood function (\ref{eqll20}) to 0, we have \cite{Xie2018}
\begin{equation}
\label{eqEstc}
\hat{s}_{k,t}=\frac{\sum_{i=k+1}^{t}(i-k)\left(\Delta_{ i}-\mu\right)}{\sum_{i=k+1}^{t}(i-k)^{2}}.
\end{equation}

Let $\tau = t-k$ be the number of samples after the change-point $k$ and $U_{k, t}=\left(A_{\tau}\right)^{-1 / 2} W_{k, t}$, 
where $A_{\tau}=\sum_{i=1}^{\tau} i^{2}$ and $W_{k, t}=\sum_{i=k+1}^{t}(i-k)\left(\Delta_{ i}-\mu \right) / \sigma_{n}$.
Substitution of (\ref{eqEstc}) into (\ref{eqll20}), 
we obtain the following GLRT procedure 
\begin{equation}
\label{eq24GLRT}
t^*= \inf \left\{t: \max _{0 \leq k<t}  \left[U_{k, t}^{2} / 2\right]  \geq h\right\},
\end{equation}
where $h$ is a prescribed threshold.
Since distribution of $\Delta_i$ under $\mathcal{H}_o$ is known or can be estimated from the measure, $h$ can be chosen based on the desired false alarm probability.


\subsubsection{Final Method}
  

At each iteration $t$, the slope change detection technique introduced in Section \ref{sec3b}, is utilized on $\bm{\Delta}^{(t)}$ to find the first abrupt change.
Stop until the first abrupt change point is detected or reaching the maximum iteration number $N-1$.
The proposed algorithm is summarized in Algorithm \ref{alg:5GP}.
\begin{algorithm}
\caption{5G Positioning and Synchronization
Algorithm}\label{alg:5GP}
\begin{algorithmic}[1]
\Require $N \geq 2$ sets of channel parameters ($\tau_j < \tau_{j+1}$)

\If {$N=2$}
\State  Estimate $\bm{p}_r$ from the $k=2$ paths using (\ref{eq_esty}).
\Else
\For{$t = 1$}
 \State  Estimate $\bm{\mu}_r^{(1)}$ from the first $k=2$ paths by (\ref{eq_esty}).
\EndFor

 \State $t = t + 1$
        \State  Estimate $\bm{\mu}_r^{(t)}$ from the first $k+t-1$ paths by (\ref{eq_esty}).
        
        \State  Compute $\Delta_{t}$ using (\ref{eq16delta}).
        
        \State Construct $\bm{\Delta}^{(t)}$ using (\ref{eq17ot})
 
 \State Estimate $t^*$ using (\ref{eq24GLRT}).
 



 \If{$\exists t^*$}
\State  Estimate $\bm{p}_r$ using the selected $t^*$ paths.
\Else
\State Estimate $\bm{p}_r$ using all the paths.
\EndIf

\EndIf 

\end{algorithmic}
\end{algorithm}


\section{Numerical Results}\label{sec4}

In this section, we evaluate the performance of the proposed method based on realistic ray-tracing data. 
\subsection{Simulation Scenario}
\begin{figure}
\centering
\includegraphics[width=0.425\textwidth]{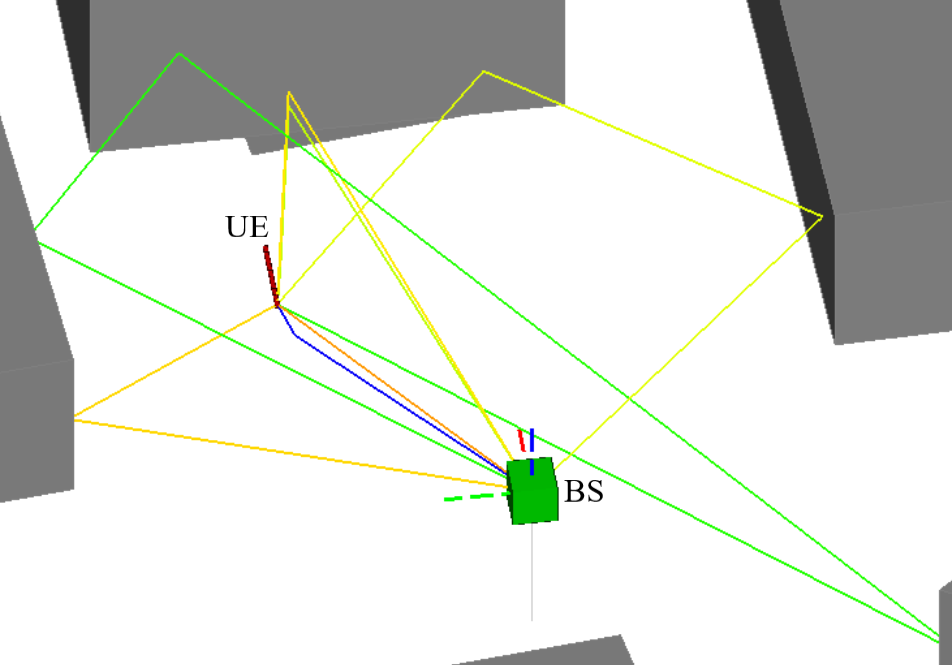}
\caption{Wireless Insite simulation setup, including the BS, the UE and several objects, which reflect and scatter the signal.}
\label{figSce}
\end{figure}

In the following simulations,
3D Wireless Prediction Software Wireless InSite  is utilized to generate the channel measurements. 
It is a suite of ray-tracing models and high-fidelity EM solvers for the analysis of site-specific radio wave propagation and wireless communication systems.
The BS is located at $\bm{p}_{t} = [621, 447, 30]^{\mathrm{T}}$, and 10 different UE positions are considered, for the $i$th UE position 
\begin{equation}
\bm{p}_{r_i} = [600, 499+i, 1.5]^{\mathrm{T}}, \text{ where }i = 1, 2, \cdots, 10.
\end{equation}
The clock bias is set to $\tau_B = 330$ ns. 
Gaussian noises are added on the path parameters,
$\mathcal{N}(0,\sigma_a)$ for AOA and AOD measurements, and $\mathcal{N}(0,\sigma_r)$ for TOA measurements.
As shown in Fig.~\ref{figSce}, a complex urban and mixed path environment is considered. Based on this environment, the ray-tracer determines all feasible propagation paths and returns their channel gain  $\gamma_{n}\ge,  0$,  AOD $(\phi_{d,n}, \theta_{d,n})$, AOA  $(\phi_{a,n}, \theta_{a,n})$, and  propagation distance $d_n$. Different levels of measurement error are added, as will be explained shortly. 

The performance of the method is evaluated in terms of two performance metrics: positioning root-mean-square error (RMSE) (\ref{eq19rmsep}) and clock bias RMSE (\ref{eq20rmseb}), which are given by
\begin{align}
\label{eq19rmsep}
    \text{RMSE}_{\bm{p}_r} &= \sqrt \frac{\sum_{k=1}^{K}\|\hat{\bm{p}}_{r,k} - \bm{p}_r \|}{K}, \\
    \label{eq20rmseb}
    \text{RMSE}_{\tau_B} &= \sqrt \frac{\sum_{k=1}^{K}\| \hat{\tau}_{B_k} - \tau_B \|}{K},
\end{align}
where $K=500$ is the number of independent runs, $\hat{\bm{p}}_{r,k}$ and $\hat{\tau}_{B_k}$ are the estimated UE position and clock bias for the $k$th trial, respectively.  
As a benchmark, the proposed method is compared with using all the paths (which we expect will degrade performance) and only using the single bounce paths (which is an optimistic performance bound). 

\subsection{Results and Discussion}

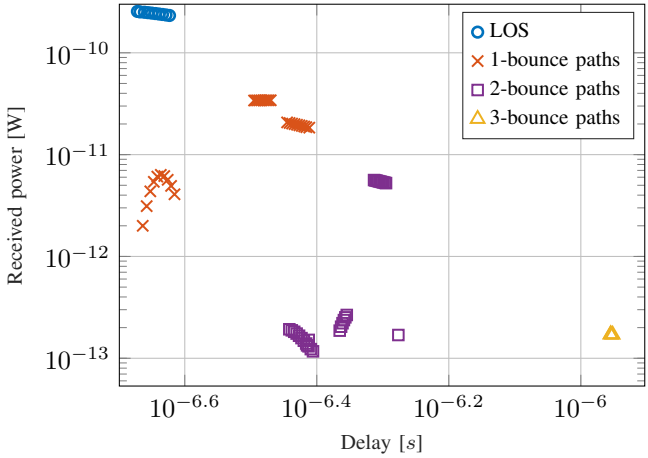
\begin{figure}
\centering
%
%
\definecolor{mycolor1}{rgb}{0.00000,0.44700,0.74100}%
\definecolor{mycolor2}{rgb}{0.85000,0.32500,0.09800}%
\definecolor{mycolor3}{rgb}{0.92900,0.69400,0.12500}%
\definecolor{mycolor4}{rgb}{0.49400,0.18400,0.55600}%
\begin{tikzpicture}

\begin{axis}[%
width=2.75in,
height=2in,
at={(0.758in,0.481in)},
scale only axis,
xmode=log,
xmin=2e-07,
xmax=1.25e-06,
xminorticks=true,
xlabel style={font=\color{white!15!black}},
xlabel={\footnotesize{Delay [$s$]}},
ymin=0,
ymax=3e-10,
ymode=log,
ylabel style={font=\color{white!15!black}},
ylabel={\footnotesize{Received power [W]}},
axis background/.style={fill=white},
xmajorgrids,
xminorgrids,
ymajorgrids,
legend style={legend cell align=left, align=left, draw=white!15!black,font=\footnotesize}
]
\addplot [color=mycolor1, line width=1.0pt, only marks, mark=o, mark options={solid, mycolor1}]
  table[row sep=crcr]{%
2.13e-07	2.55e-10\\
2.15e-07	2.52e-10\\
2.18249e-07	2.50167e-10\\
2.21062e-07	2.47856e-10\\
2.2389e-07	2.45465e-10\\
2.26731e-07	2.43008e-10\\
2.29586e-07	2.40486e-10\\
2.32454e-07	2.37919e-10\\
2.35333e-07	2.35304e-10\\
2.38225e-07	2.32648e-10\\
};
\addlegendentry{LOS}

\addplot [color=mycolor2, line width=0.75pt, only marks, mark=x, mark options={solid, mycolor2, mark size=3pt}]
  table[row sep=crcr]{%
3.2e-07	3.4e-11\\
3.88e-07	1.84e-11\\
2.17e-07	2e-12\\
3.22e-07	3.41e-11\\
3.84e-07	1.86e-11\\
2.2e-07	3.12e-12\\
3.24101e-07	3.40918e-11\\
3.81192e-07	1.8889e-11\\
2.22793e-07	4.35963e-12\\
3.26192e-07	3.41185e-11\\
3.78021e-07	1.91395e-11\\
2.2555e-07	5.40555e-12\\
3.28303e-07	3.41405e-11\\
3.74853e-07	1.93937e-11\\
2.28322e-07	6.08415e-12\\
3.30434e-07	3.4157e-11\\
3.71688e-07	1.96521e-11\\
2.31109e-07	6.32951e-12\\
3.32586e-07	3.41696e-11\\
3.68526e-07	1.9915e-11\\
2.3391e-07	6.16254e-12\\
3.34756e-07	3.41767e-11\\
3.65367e-07	2.01813e-11\\
2.36725e-07	5.66044e-12\\
3.36946e-07	3.41798e-11\\
3.62211e-07	2.04527e-11\\
2.39553e-07	4.9289e-12\\
3.39155e-07	3.41775e-11\\
3.59058e-07	2.07276e-11\\
2.42395e-07	4.07868e-12\\
};
\addlegendentry{1-bounce paths}

\addplot [color=mycolor3, line width=0.75pt, only marks, mark=square, mark options={solid, mycolor4}]
  table[row sep=crcr]{%
5.07e-07	5.25e-12\\
3.9e-07	1.23e-13\\
3.93e-07	1.17e-13\\
5.05e-07	5.29e-12\\
3.85e-07	1.34e-13\\
3.87e-07	1.31e-13\\
5.02422e-07	5.33102e-12\\
3.87178e-07	1.51555e-13\\
3.83812e-07	1.39643e-13\\
5.00076e-07	5.37155e-12\\
5.29249e-07	1.69204e-13\\
3.80663e-07	1.48693e-13\\
4.97741e-07	5.41278e-12\\
4.31341e-07	1.86831e-13\\
3.77517e-07	1.57768e-13\\
4.95417e-07	5.45444e-12\\
4.33453e-07	2.04188e-13\\
3.74374e-07	1.6656e-13\\
4.93105e-07	5.49667e-12\\
4.35585e-07	2.21034e-13\\
3.71235e-07	1.74747e-13\\
4.90805e-07	5.53962e-12\\
4.37736e-07	2.37137e-13\\
3.68099e-07	1.82033e-13\\
4.88517e-07	5.58316e-12\\
4.39907e-07	2.52267e-13\\
3.64967e-07	1.88105e-13\\
4.86241e-07	5.6273e-12\\
4.42096e-07	2.66226e-13\\
3.61838e-07	1.92695e-13\\
};
\addlegendentry{2-bounce paths}

\addplot [color=mycolor4, line width=0.75pt, only marks, mark=triangle, mark options={solid, mycolor3, mark size=3pt}]
  table[row sep=crcr]{%
1.11e-06	1.72e-13\\
1.11e-06	1.71e-13\\
1.11744e-06	1.70455e-13\\
};
\addlegendentry{3-bounce paths}

\end{axis}
 
\end{tikzpicture}%
\caption{The path information obtained from Wireless Insite software for 10 different scenarios.}
\label{figSim}
\end{figure}

Fig.~\ref{figSim} shows, for each UE position, the amplitude of the paths as a function of delay for the different UE locations. LOS, single and multiple bounce paths are observable as shown by the different colors. We observe that LOS paths arrive first and have largest power. Generally single-bounce paths arrive before multi-bounce paths and have a larger power. However, there are cases where multi-bounce paths arrive with greater power than single-bounce paths. 

\begin{figure}
\centering
%
%
\definecolor{mycolor1}{rgb}{0.00000,0.44700,0.74100}%
\definecolor{mycolor2}{rgb}{0.85000,0.32500,0.09800}%
\definecolor{mycolor3}{rgb}{0.92900,0.69400,0.12500}%
\definecolor{mycolor4}{rgb}{0.49400,0.18400,0.55600}%
\definecolor{mycolor5}{rgb}{0.46600,0.67400,0.18800}%
\definecolor{mycolor6}{rgb}{0.30100,0.74500,0.93300}%
\definecolor{mycolor7}{rgb}{0.63500,0.07800,0.18400}%
\begin{tikzpicture}

\begin{axis}[%
width=2.75in,
height=2.15in,
at={(0.758in,0.484in)},
scale only axis,
xmode=log,
xmin=0.001,
xmax=0.1,
xminorticks=true,
xlabel style={font=\color{white!15!black}},
xlabel={\footnotesize{Standard deviation of angle measurements $\sigma_a$ [rad]}},
ymode=log,
ymin=0.0978710405300056,
ymax=3000,
yminorticks=true,
ylabel style={font=\color{white!15!black}},
ylabel={\footnotesize{Position RMSE [m]}},
axis background/.style={fill=white},
xmajorgrids,
xminorgrids,
ymajorgrids,
yminorgrids,
legend columns=1,
legend style={at={(0.025,0.45)}, anchor=south west, legend cell align=left, align=left, draw=white!15!black,font=\footnotesize}
]
\addplot [color=mycolor1, dash dot, line width=1.0pt]
  table[row sep=crcr]{%
0.001	121.676315105559\\
0.002	120.224619455182\\
0.005	116.391049643658\\
0.01	114.225455119811\\
0.02	113.050534231859\\
0.05	111.349335429393\\
0.1	110.40840253386\\
};
\addlegendentry{All paths, $\sigma{}_\text{r} \in \{2, 1, 0.1\}$  m}

\addplot [color=mycolor2, line width=1.0pt, only marks, mark=triangle, mark options={solid, rotate=180, mycolor2, mark size=3pt}]
  table[row sep=crcr]{%
0.001	4.1869769002902\\
0.002	4.14496526721809\\
0.005	4.33398330772117\\
0.01	4.90313629210686\\
0.02	6.13501910953384\\
0.05	12.1781651106674\\
0.1	24.1187715208829\\
};
\addlegendentry{Proposed, $\sigma{}_\text{r}$ = 2 m}

\addplot [color=mycolor3, line width=1.0pt, mark options={solid, mycolor3}]
  table[row sep=crcr]{%
0.001	4.17460850362621\\
0.002	4.13224708174817\\
0.005	4.29894574360405\\
0.01	4.89153035045575\\
0.02	6.08093548223337\\
0.05	12.0960408280411\\
0.1	24.1340256268959\\
};
\addlegendentry{Single-bounce, $\sigma{}_\text{r}$ = 2 m}


\addplot [color=mycolor4, line width=1.0pt, only marks, mark=square, mark options={solid, mycolor4}]
  table[row sep=crcr]{%
0.001	2.07068362825728\\
0.002	2.27657838398792\\
0.005	2.41178811573683\\
0.01	3.08563668639204\\
0.02	4.82025399917982\\
0.05	10.9177535715813\\
0.1	23.2660649371507\\
};
\addlegendentry{Proposed, $\sigma{}_\text{r}$ = 1 m}

\addplot [color=mycolor2, line width=1.2pt, dotted]
  table[row sep=crcr]{%
0.001	2.07068362825728\\
0.002	2.27657838398792\\
0.005	2.41178811573683\\
0.01	3.07409777272994\\
0.02	4.79733257886871\\
0.05	10.9087383477338\\
0.1	23.2686545188615\\
};
\addlegendentry{Single-bounce, $\sigma{}_\text{r}$ = 1 m}


\addplot [color=mycolor1, line width=1.0pt, only marks, mark=*, mark options={solid, mycolor6}]
  table[row sep=crcr]{%
0.001	0.3071648066451\\
0.002	0.497517631071924\\
0.005	1.08060029377323\\
0.01	2.18168588624767\\
0.02	4.38517392599271\\
0.05	11.064752076789\\
0.1	24.0099413364541\\
};
\addlegendentry{Proposed, $\sigma{}_\text{r}$ = 0.1 m}

\addplot [color=mycolor7, dashed, line width=1.0pt]
  table[row sep=crcr]{%
0.001	0.3071648066451\\
0.002	0.497517631071924\\
0.005	1.08060029377323\\
0.01	2.18168588624767\\
0.02	4.35223284440675\\
0.05	11.0117411565673\\
0.1	23.9835003651633\\
};
\addlegendentry{Single-bounce, $\sigma{}_\text{r}$ = 0.1 m}

\end{axis}

\end{tikzpicture}%
\caption{Gaussian noise $\mathcal{N}(0,\sigma_a)$ is added on the AOA, AOD measurements, and $\mathcal{N}(0,\sigma_r)$ is added on TOA measurements.}
\label{fig4}
\end{figure}

\begin{figure}
\centering
%
%
\definecolor{mycolor1}{rgb}{0.00000,0.44700,0.74100}%
\definecolor{mycolor2}{rgb}{0.85000,0.32500,0.09800}%
\definecolor{mycolor3}{rgb}{0.92900,0.69400,0.12500}%
\definecolor{mycolor4}{rgb}{0.49400,0.18400,0.55600}%
\definecolor{mycolor5}{rgb}{0.46600,0.67400,0.18800}%
\definecolor{mycolor6}{rgb}{0.30100,0.74500,0.93300}%
\definecolor{mycolor7}{rgb}{0.63500,0.07800,0.18400}%
\begin{tikzpicture}

\begin{axis}[%
width=2.75in,
height=2.15in,
at={(0.758in,0.484in)},
scale only axis,
xmode=log,
xminorticks=true,
xmin=0.001,
xmax=0.1,
xlabel style={font=\color{white!15!black}},
xlabel={\footnotesize{Standard deviation of angle measurements $\sigma_a$ [rad]}},
ymode=log,
ymin=1,
ymax=1000,
yminorticks=true,
ylabel style={font=\color{white!15!black}},
ylabel={\footnotesize{Clock bias RMSE [ns]}},
axis background/.style={fill=white},
xmajorgrids,
xminorgrids,
ymajorgrids,
yminorgrids,
legend style={at={(0.025,0.475)}, anchor=south west, legend cell align=left, align=left, draw=white!15!black,font=\footnotesize}
]
\addplot [color=mycolor1, dash dot, line width=1.0pt]
  table[row sep=crcr]{%
0.001	750.54329346632\\
0.002	745.544165391413\\
0.005	731.318104050541\\
0.01	723.118252416974\\
0.02	719.444765661545\\
0.05	710.496736835095\\
0.1	    705.868470015349\\
};
\addlegendentry{All paths, $\sigma{}_\text{r} \in \{2, 1, 0.1\}$  m}

\addplot [color=mycolor2, line width=1.2pt, only marks, mark=triangle, mark options={solid, rotate=180, mycolor2, mark size=3pt}]
  table[row sep=crcr]{%
0.001	18.7321372620337\\
0.002	18.5293943053319\\
0.005	19.1026460436967\\
0.01	20.5645610748945\\
0.02	23.8027241671994\\
0.05	43.6288697897518\\
0.1	85.0026965013924\\
};
\addlegendentry{Proposed, $\sigma{}_\text{r}$ = 2 m}

\addplot [color=mycolor3, line width=1pt]
  table[row sep=crcr]{%
0.001	18.7212123307325\\
0.002	18.5213682306527\\
0.005	19.0920738728331\\
0.01	20.5473254358514\\
0.02	23.8016560410299\\
0.05	43.4413225062587\\
0.1	85.1052502024856\\
};
\addlegendentry{Single-bounce, $\sigma{}_\text{r}$ = 2 m}


\addplot [color=mycolor5, line width=1.0pt, only marks, mark=square, mark options={solid, mycolor5, mark size=2pt}]
  table[row sep=crcr]{%
0.001	9.34608136054803\\
0.002	9.93922210079851\\
0.005	10.1192514460845\\
0.01	12.1707771868024\\
0.02	17.3285909661601\\
0.05	37.5641428044841\\
0.1	80.4836310495566\\
};
\addlegendentry{Proposed, $\sigma{}_\text{r}$ = 1 m}

\addplot [color=mycolor6, line width=1.5pt, dotted]
  table[row sep=crcr]{%
0.001	9.34608136054803\\
0.002	9.93922210079851\\
0.005	10.1192514460845\\
0.01	12.1485262335528\\
0.02	17.3077306974028\\
0.05	37.574373911269\\
0.1	80.5491998573095\\
};
\addlegendentry{Single-bounce, $\sigma{}_\text{r}$ = 1 m}


\addplot [color=mycolor1, line width=1.2pt, only marks, mark=*, mark options={solid, mycolor1}]
  table[row sep=crcr]{%
0.001	1.17708834252642\\
0.002	1.78933355371682\\
0.005	3.65397069969353\\
0.01	7.42487956792739\\
0.02	14.6345750538175\\
0.05	37.9262066184826\\
0.1	84.2784019259075\\
};
\addlegendentry{Proposed, $\sigma{}_\text{r}$ = 0.1 m}

\addplot [color=mycolor2, dashed, line width=1.2pt]
  table[row sep=crcr]{%
0.001	1.17708834252642\\
0.002	1.78933355371682\\
0.005	3.65397069969353\\
0.01	7.42487956792739\\
0.02	14.6188110239259\\
0.05	37.8760696970185\\
0.1	84.1887915240761\\
};
\addlegendentry{Single-bounce, $\sigma{}_\text{r}$ = 0.1 m}

\end{axis}

\end{tikzpicture}%
\caption{Gaussian noise $\mathcal{N}(0,\sigma_a)$ is added on the AOA, AOD measurements, and $\mathcal{N}(0,\sigma_r)$ is added on TOA measurements.}
\label{fig5}
\end{figure}

 The performance is evaluated by considering 10 different scenarios as shown in Fig.
\ref{figSim}, as well as considering different AOA, AOD and TOA measurement errors. 
The positioning and synchronization performance is shown in Fig.~\ref{fig4} and Fig.
\ref{fig5}, respectively, as a function of the AOA and AOD error standard deviation, for different levels of TOA standard deviation (expressed in meters). 
It can be observed that sub-meter accuracy is achievable when the angle error standard deviation is small (below 0.01 rad) and the TOA error standard deviation is around 0.1 m. The proposed method performs robustly,  even in the presence of  multi-bounce paths, attaining the performance of using only single bounce paths. With the increase of AOA and AOD measurement errors,  positioning RMSE of all methods increase, but  is still better than that of using all paths. This shows that the proposed method can distinguish single- and multi-bounce paths in multipath environments and can control errors to a small level. In terms of clock bias estimation performance, similar conclusions can be drawn.

\section{Conclusion}\label{sec5}
We propose a robust algorithm to mitigate the effect of multi-bounce paths, based on a combination of weighted least squares and a change detection approach. Numerical results are provided to evaluate the performance of the algorithm, and the results show that it can greatly improve the positioning accuracy.  
One of the assumptions we made is that the first two arrival paths are single bounce paths, which may not always be true. 
In future work, we will further improve the applicability of the algorithm as well as the localization accuracy.

\balance
\bibliographystyle{IEEEtran}
\bibliography{bib/References.bib}

\end{document}